\documentclass[10pt,conference,letterpaper]{IEEEtran}
\IEEEoverridecommandlockouts

\usepackage{cite}
\usepackage{amsmath,amssymb,amsfonts}
\usepackage{graphicx}
\usepackage{textcomp}
\usepackage{xcolor}
\def\BibTeX{{\rm B\kern-.05em{\sc i\kern-.025em b}\kern-.08em
    T\kern-.1667em\lower.7ex\hbox{E}\kern-.125emX}}

\usepackage{booktabs}
\usepackage{subfig}
\usepackage{array}
\newcolumntype{L}{>{\centering\arraybackslash}m{0.4in}}
\usepackage{caption}
\usepackage{wrapfig}
\usepackage{algpseudocode}
\algrenewcommand{\algorithmiccomment}[1]{\hfill\textbf{//}\,#1}
\usepackage{algorithm}
\algrenewcommand\alglinenumber[1]{\small #1:}
\usepackage{circledsteps}
\pgfkeys{/csteps/inner color=white}
\pgfkeys{/csteps/fill color=black}
\usepackage{pifont}
\usepackage{multirow}
\usepackage{mdframed}

\usepackage{soul}

\edef\oldtt{\ttdefault}
\usepackage[scaled]{beramono}
\usepackage[T1]{fontenc}
\renewcommand{\ttdefault}{\oldtt}
\newcommand{\bera}[1]{\text{\fontfamily{fvm}\selectfont #1}}
\usepackage{mathtools}

\usepackage{eso-pic}

\begin{document}

\AddToShipoutPictureFG*{%
  \AtPageUpperLeft{%
    \makebox[\pdfpagewidth]{\raisebox{\dimexpr-\height-20pt}{%
      \large To appear in 2025 IEEE Cross-disciplinary Conference on Memory-Centric Computing (CCMCC).
    }}%
  }%
}

\title{CryptoSRAM: Enabling High-Throughput Cryptography on MCUs via In-SRAM Computing
}

\author{\IEEEauthorblockN{Jingyao Zhang}
\IEEEauthorblockA{\textit{Department of Computer Science and Engineering} \\
\textit{University of California, Riverside}\\
jzhan502@ucr.edu}
\and
\IEEEauthorblockN{Elaheh Sadredini}
\IEEEauthorblockA{\textit{Department of Computer Science and Engineering} \\
\textit{University of California, Riverside}\\
elaheh@cs.ucr.edu}
}

\maketitle

\begin{abstract}
Secure communication is a critical requirement for Internet of Things (IoT) devices, which are often based on Microcontroller Units (MCUs). Current cryptographic solutions, which rely on software libraries or dedicated hardware accelerators, are fundamentally limited by the performance and energy costs of data movement between memory and processing units. This paper introduces CryptoSRAM, an in-SRAM computing architecture that performs cryptographic operations directly within the MCU's standard SRAM array. By repurposing the memory array into a massively parallel processing fabric, CryptoSRAM eliminates the data movement bottleneck. This approach is well-suited to MCUs, which utilize physical addressing and Direct Memory Access (DMA) to manage SRAM, allowing for seamless integration with minimal hardware overhead. Our analysis shows that for common cryptographic kernels, CryptoSRAM achieves throughput improvements of up to 74$\times$ and 67$\times$ for AES and SHA3, respectively, compared to a software implementation. Furthermore, our solution delivers up to 6$\times$ higher throughput than existing hardware accelerators for AES. CryptoSRAM demonstrates a viable and efficient architecture for secure communication in next-generation IoT systems.
\end{abstract}


\section{Introduction}

\begin{figure}[htp]
\centerline{\includegraphics[width=3.4in]{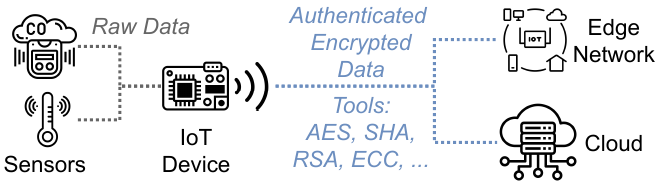}}
\caption{A typical secure communication workflow in an IoT device. On-chip cryptographic modules are used to ensure data confidentiality and integrity.}
\label{intro}
\end{figure}

The proliferation of Internet of Things (IoT) devices across consumer, commercial, and industrial domains has underscored the critical need for secure and efficient data communication \cite{madakam2015internet, farooq2015review}. Given the sensitive nature of the data collected and transmitted by these devices, cryptographic operations are fundamental for ensuring data confidentiality and integrity \cite{vcolakovic2018internet, xu2014security, yadav2018iot}. As illustrated in Figure \ref{intro}, standard security protocols in Microcontroller Unit (MCU) based IoT devices rely on cryptographic primitives such as the Advanced Encryption Standard (AES) and Secure Hash Algorithm 3 (SHA3) to protect data before transmission \cite{daemen2001reijndael, dworkin_sha-3_2015}. Similar concerns arise in other resource-constrained domains \cite{Kazemi2018-dm,Murshed2022-id,Xiao2019-mj,Li2023-lv,Li2023-uv,Jia2024-de}.

Conventional approaches to implementing these cryptographic tasks on MCUs typically involve dedicated hardware accelerators \cite{zhang2016compact, martino2020designing, lara2020lightweight}. While functional, this paradigm introduces significant performance and energy-efficiency bottlenecks. The inefficiency stems primarily from the substantial data movement required between on-chip SRAM, the CPU, and the separate cryptographic hardware engines \cite{suarez-albelaPracticalEvaluationRSA2018, suarez-albelaPracticalPerformanceComparison2018, munozAnalyzingResourceUtilization2018, biasizzoHardwareImplementationAES2005, contiIoTEndpointSystemonchip2017}. This data-transfer overhead not only consumes considerable energy but also constrains the overall throughput of secure communications. While algorithmic optimizations such as lightweight cryptography \cite{dhandaLightweightCryptographySolution2020, yallaLightweightCryptographyFPGAs2009, ghoshLightweightPostQuantumSecureDigital2019, mustafaLightweightPostQuantumLatticeBased2020, saarinenRingLWECiphertextCompression2017} and efficient data compression \cite{azarEnergyEfficientIoT2019, stojkoskaDataCompressionEnergy2017, alarifiNovelHybridCryptosystem2020} aim to reduce computational complexity, improving the underlying hardware's efficiency remains a paramount challenge.

To address this data movement problem, we explore the paradigm of processing-in-memory (PIM), which collocates computation with data storage, whether in SRAM \cite{aga2017compute, yinXNORSRAMInMemoryComputing2020a, fujikiDualityCacheData2019g, jhangChallengesTrendsSRAMBased2021, liuNSCIMCurrentModeComputationinMemory2020, Zhang2023-rt,Sadredini2020-ha,Angstadt2018-sl,Sadredini2021-hy}, DRAM \cite{ahnScalableProcessinginmemoryAccelerator2015b, seshadriAmbitInmemoryAccelerator2017b, kimGradPIMPracticalProcessinginDRAM2021, hajinazarSIMDRAMFrameworkBitserial2021a, Ghinani2025-pf,Lenjani2020-mg}, or storage-class memory \cite{kooSummarizerTradingCommunication2017, parkFlashCosmosInFlashBulk2022, parkInStorageComputingHadoop2016, ruanINSIDERDesigningInStorage, tsengMorpheusCreatingApplication2016}, to minimize data transfer. 
Among these PIM approaches, in-SRAM computing presents unique advantages for MCU-based cryptographic acceleration. Unlike DRAM-based PIM, which requires complex refresh cycles and exhibits higher access latencies, SRAM offers deterministic, single-cycle access crucial for real-time cryptographic operations. Furthermore, SRAM's inherent array structure enables fine-grained bit-level manipulation through repurposed peripheral circuits, making it naturally suited for the bit-oriented operations prevalent in cryptographic algorithms. In-SRAM computing transforms traditional SRAM arrays into massively parallel computational fabrics by leveraging the analog properties of memory cells. By activating multiple wordlines simultaneously and repurposing sense amplifiers as logic gates, these architectures can perform Boolean operations across entire rows of data in a single cycle \cite{aga2017compute, fujikiDualityCacheData2019g}. This parallelism is particularly beneficial for cryptographic workloads, where operations like substitution boxes (S-boxes), permutations, and bitwise XOR operations dominate the computational profile.

Moreover, the integration of in-SRAM computing within MCU architectures is particularly feasible due to several architectural synergies. First, MCUs already contain substantial on-chip SRAM that serves as the primary data memory, eliminating the need for additional memory resources. Second, the typical memory capacities in MCUs (ranging from tens to hundreds of kilobytes) align well with the data footprints of cryptographic operations, allowing entire encryption blocks and intermediate states to reside within the computational fabric. Third, the relatively simple memory hierarchies in MCUs—often lacking complex cache coherence protocols—simplify the integration of compute-capable memory arrays without introducing consistency challenges.

In this work, we propose \textit{CryptoSRAM}, a specialized in-SRAM computing architecture tailored for cryptographic acceleration within MCUs. Our core insight is that the architectural features of MCUs—specifically, the use of a physical memory address space and the availability of Direct Memory Access (DMA) for memory management—make the integration of in-SRAM computing particularly feasible with minimal system-level modifications (detailed in Section \ref{sec:feasibility}). By performing cryptographic operations directly within the SRAM array, CryptoSRAM effectively eliminates the data movement between memory, CPU, and accelerators.


The proposed architecture yields substantial performance and efficiency gains. Our evaluation demonstrates throughput improvements of up to 74× for AES and 67× for SHA3 compared to standard software implementations. Furthermore, CryptoSRAM surpasses existing MCU hardware accelerators by up to 6× in throughput for AES. Critically for battery-powered IoT devices, our approach extends operational lifetime up to 283 hours for continuous secure communication, even allowing for the main processor to be power-gated. These benefits are a direct result of the massive parallelism inherent in the in-SRAM computing fabric and the near-elimination of data transfer overhead.

The primary contributions of this paper are:
\begin{itemize}
\item The design and integration of an in-SRAM computing architecture, \textit{CryptoSRAM}, specifically for MCUs. We demonstrate its feasibility by leveraging native MCU features like physical addressing and DMA to minimize system overhead (Section \ref{implementation}).
\item An algorithm-architecture co-design that maps cryptographic primitives (AES and SHA3) onto the SRAM fabric. We employ bit-slicing and a novel \textit{lane-per-row} data mapping strategy to maximize the inherent parallelism of the memory array, significantly accelerating computation (Section \ref{alg-co-design}).
\item A comprehensive evaluation of CryptoSRAM, quantifying its performance and energy efficiency against conventional software and hardware-accelerated solutions. The results confirm significant throughput gains and a dramatic extension of device battery life, highlighting its suitability for power-constrained IoT applications (Section \ref{evaluation}).
\end{itemize}

The remainder of this paper is organized as follows. Section \ref{background} presents background on secure communication in MCUs. Section \ref{sec:feasibility} discusses the feasibility of integrating in-SRAM computing within existing MCU systems. Section \ref{implementation} details the proposed architecture. Section \ref{alg-co-design} describes the algorithm-architecture co-design methodology. Section \ref{evaluation} presents our evaluation results. Section \ref{relatedwork} reviews related work, and Section \ref{conclusion} concludes the paper.

\section{Background and Motivation}\label{background}

\subsection{Microcontroller System Architecture}

\begin{figure}[!ht]
\centerline{\includegraphics[width=2.8in]{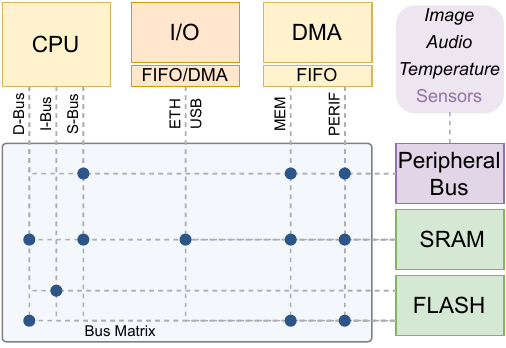}}
\caption{Diagram of microcontroller system architecture.
}
\label{mcu-arch}
\end{figure}

Microcontrollers (MCUs), due to their power efficiency, compact size, and cost-effectiveness, have found extensive applications in numerous fields, including image sensing, audio sensing, temperature sensing, and many more, thus becoming the cornerstone of the IoT ecosystem. 
A typical MCU system \cite{UnknownUnknown-tc, UnknownUnknown-az, UnknownUnknown-hy, UnknownUnknown-ts, UnknownUnknown-ka,STM32DMACheat} comprises a variety of elements, such as the Bus Matrix, which regulates communication among various modules in the system, a CPU, I/O devices (e.g., Ethernet and USB), the Peripheral Bus for connecting multiple sensors, a memory system, and Direct Memory Access (DMA). The memory system primarily consists of SRAM and FLASH memory. Given its non-volatile nature, FLASH memory is predominantly used for storing program codes and data that need to be preserved following a power outage. As shown in Figure \ref{mcu-arch}, the CPU's instruction bus (I-Bus) is solely connected to the FLASH memory, serving to retrieve program codes stored therein.

SRAM, on the other hand, owing to its faster read-write speeds, typically functions as temporary storage for the processor and is used to store variables from currently running programs. To enhance performance, alleviate the workload on the CPU, and accelerate data transmission rates, a typical MCU system includes one or more DMA \cite{STM32DMACheat}. This allows peripheral and I/O devices to access the SRAM directly, bypassing the CPU. Some DMAs even support SRAM-to-SRAM data movement \cite{STM32DMACheat}.
Within the MCU environment, DMAs are commonly utilized to handle tasks requiring substantial data transfers, such as audio/video processing, network communication, analog-to-digital conversions, and so forth. The cohesive integration of these components forms the backbone of the MCU system architecture, underscoring its role in the broader IoT ecosystem.

\subsection{Motivation: Supply/Demand Analysis}


In the field of IoT devices, such as cameras and microphones, the use of MCU systems is prevalent. These devices frequently process and transmit sensitive data, such as personal video and audio information, which necessitates stringent measures for preserving privacy and ensuring data integrity.

In this context, secure communication is paramount, as it provides a channel for data encryption and decryption during transmission. This guards against unauthorized access and tampering. To establish this secure communication, cryptographic tools like AES and SHA3 are enlisted. These industry-standard encryption and hash algorithms play an indispensable role in securing IoT communications.

In this section, we delve into a representative application scenario of an MCU system, focusing on image and audio sensing in IoT devices. We consider a typical setup that includes an STM32L562 MCU system \cite{STM32L5x2ArmCortex}, an OV7670 image sensor \cite{OV7670DatasheetOmniVision}, and an MP34DT01-M audio sensor \cite{MEMSAudioSensor}. The motivation stems from the realization that, despite the low-end sensors utilized in this case, the demand for secure communication throughput significantly surpasses the system's capacity. This imbalance becomes even more pronounced when employing higher-end or more numerous sensors.

\begin{figure}[ht]
\centerline{\includegraphics[width=3in]{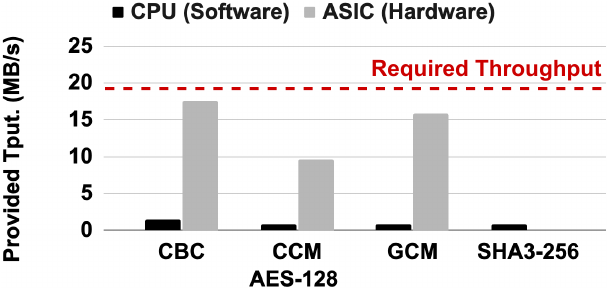}}
\caption{Provided and required throughput of secure processing and transmission on an image and audio sensing system (STM32L562 \cite{STM32L5x2ArmCortex} + OV7670 \cite{OV7670DatasheetOmniVision} + MP34DT01-M \cite{MEMSAudioSensor}).}
\label{motivation}
\end{figure}

As per the datasheet, when operating at VGA resolution with a YUV422 pixel format at 30 frames per second, the OV7670 demands a throughput of 18.43MB/s \cite{OV7670DatasheetOmniVision}. The MP34DT01-M, in its typical mode, necessitates a throughput of 0.3MB/s \cite{MEMSAudioSensor}. Consequently, the combined required throughput stands at 18.73MB/s, which exceeds the capabilities of the MCU system, as shown in Figure \ref{motivation}.

In Figure \ref{motivation}, we illustrate the system's performance under different cryptographic operations, including AES128-CBC, AES128-CCM, AES128-GCM, and SHA3-256. 
AES is a symmetric encryption method, and Cipher Block Chaining (CBC) is an operation mode which uses previous block's cipher as an input to the next block for added security. Counter with CBC-Message Authentication Code (CCM) combines CBC for encryption and a form of counter mode for authentication, providing both confidentiality and data integrity. Galois/Counter Mode (GCM) is an operation mode for AES that offers both encryption and authentication using a single key. GCM is known for its high performance due to its support for parallel processing. Secure Hash Algorithm 3 (SHA3) with 256-bit output is a cryptographic hash function. It provides a one-way function to verify data integrity.

In Figure \ref{motivation}, we highlight a comparison between the performance outcomes when using a CPU software approach versus an ASIC hardware approach. Noticeably, ASIC in STM32L562 \cite{STM32L5x2ArmCortex} does not natively support SHA3-256, therefore its data is absent.
Figure \ref{motivation} emphasizes that the present system struggles to meet the total required throughput for image and audio data processing. To address this challenge, we need to enhance the system's throughput while ensuring low overhead and energy consumption. Integrating in-SRAM computing into the MCU system can provide a viable solution, taking advantage of its high parallelism and energy efficiency, to meet these demanding requirements.

\subsection{The Datapath of Sensor Data} 

\begin{figure*}[!ht]
\centerline{\includegraphics[width=6.5in]{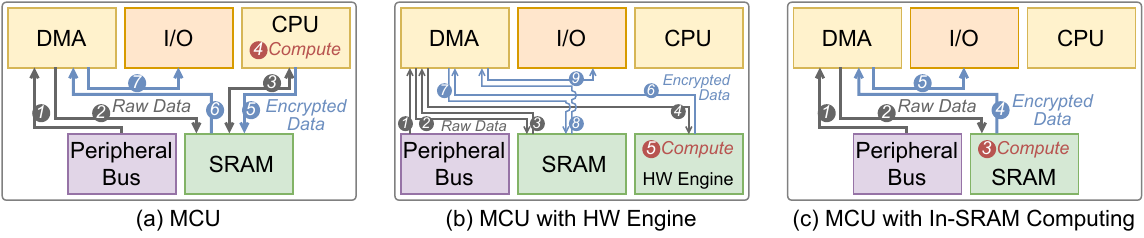}}
\caption{Datapath of sensor data in (a) conventional microcontroller, (b) conventional microcontroller with hardware security engine, and (c) Our approach, which is microcontrollers with in-SRAM computing (\textit{CryptoSRAM}).
In traditional MCU systems, 7 steps are required for the processing and secure transmission of sensor data. Although an MCU system equipped with a hardware security engine can execute cryptographic computations more rapidly, it necessitates 11 steps for ping-pong data movement between modules.
Conversely, an MCU system with in-SRAM computing, capable of performing computations directly within the SRAM, only requires 5 steps to complete the processing and secure transmission of sensor data.}
\label{design-comparison}
\end{figure*}


In this section, we aim to clarify the distinct processing and transmission paths secure sensor data takes in different architectures of MCU systems. We start with a conventional MCU architecture, move on to an MCU equipped with a hardware security engine, and then introduce the advantages offered by an MCU system that utilizes in-SRAM computing.

\textbf{Conventional MCU System:} In conventional MCU architecture, the processing of secure sensor data and its transmission follow a well-defined path. To understand this, let's take an example where data is generated by a peripheral sensor, as shown in Figure \ref{design-comparison}(a). 
Step \Circled{\scriptsize{\textit{\textbf{1}}}}: Raw data is generated by a peripheral sensor. Step \Circled{\scriptsize{\textit{\textbf{2}}}}: This raw data is then transferred into the Static Random-Access Memory (SRAM) of the MCU. Step \Circled{\scriptsize{\textit{\textbf{3}}}}: The Central Processing Unit (CPU) retrieves this raw data from the SRAM for processing. Step \Circled{\scriptsize{\textit{\textbf{4}}}}: The CPU processes the raw data, which includes operations like encryption and hashing, transforming it into encrypted data. Step \Circled{\scriptsize{\textit{\textbf{5}}}}: After processing, the CPU returns the encrypted data back to the SRAM. Step \Circled{\scriptsize{\textit{\textbf{6}}}}: From the SRAM, the encrypted data is ready for transmission and fetched by DMA. Step \Circled{\scriptsize{\textit{\textbf{7}}}}: Finally, the encrypted data is transmitted to other devices via Ethernet, typically through a protocol stack that ensures reliable delivery.
Throughout these operations, the CPU orchestrates the data flow between the peripheral sensors, SRAM, and Ethernet \cite{STM32L5x2ArmCortex,LPC4357FET256ArmCortexM4}.

\textbf{MCU System with Hardware Security Engine:} To enhance the security and efficiency of this process, MCUs often integrate hardware security engines into their architecture. These dedicated hardware units are designed to perform encryption tasks, offloading these computationally intensive operations from the CPU. They are connected as a peripheral to the MCU \cite{LPC43S70ArmCortex}, as shown in Figure \ref{design-comparison}(b). The datapath of sensor data in MCU with hardware security engine is as follows.
Step \Circled{\scriptsize{\textit{\textbf{1}}}}: Raw data is generated by a peripheral sensor.
Step \Circled{\scriptsize{\textit{\textbf{2}}}}: This raw data is then transferred from the sensor into the SRAM of the system via DMA.
Step \Circled{\scriptsize{\textit{\textbf{3}}}}: The raw data in the SRAM is prepared to be transferred to the hardware security engine (HW Engine), via DMA.
Step \Circled{\scriptsize{\textit{\textbf{4}}}}: The DMA transfers the raw data from the SRAM to the hardware security engine.
Step \Circled{\scriptsize{\textit{\textbf{5}}}}: The hardware security engine processes the raw data by performing operations such as encryption and hash value generation, transforming it into encrypted data.
Step \Circled{\scriptsize{\textit{\textbf{6}}}}: The hardware security engine then transfers the encrypted data to DMA.
Step \Circled{\scriptsize{\textit{\textbf{7}}}}: 
DMA transfers the encrypted data from the hardware security engine back to the SRAM.
Step \Circled{\scriptsize{\textit{\textbf{8}}}}: When transmission is needed, the DMA prepares to transfer the encrypted data from the SRAM to the I/O device (e.g., Ethernet or USB).
Step \Circled{\scriptsize{\textit{\textbf{9}}}}: Finally, the DMA transfers the encrypted data from the SRAM to the I/O device, and it is then transmitted to other devices.
Though a hardware security engine can work in parallel with the CPU, accelerating the encryption process and thus, the overall data transmission process, it falls short in several aspects. It struggles to cater to the computation demands from multiple peripherals, requires DMA assistance to move data from SRAM to peripherals, consumes a high amount of energy, and is limited in scalability due to the throughput capacity of DMA \cite{munozAnalyzingResourceUtilization2018,biasizzoHardwareImplementationAES2005,contiIoTEndpointSystemonchip2017}.

\textbf{MCU System with In-SRAM Computing:} The datapath of sensor data in an MCU system with in-SRAM computing, as illustrated in Figure \ref{design-comparison}(c). Step \Circled{\scriptsize{\textit{\textbf{1}}}}: Raw data is generated by the peripheral sensor, which is then captured by the DMA.
Step \Circled{\scriptsize{\textit{\textbf{2}}}}: The raw data in the DMA is stored in the SRAM.
Step \Circled{\scriptsize{\textit{\textbf{3}}}}: Thanks to the application of in-SRAM computing, the SRAM itself processes the raw data. Operations such as encryption and authentication are performed directly in the SRAM, eliminating the need for data movement to a separate processing unit.
Step \Circled{\scriptsize{\textit{\textbf{4}}}}: Upon completion of these calculations, the encrypted and authenticated data is transferred directly to the DMA.
Step \Circled{\scriptsize{\textit{\textbf{5}}}}: Finally, the I/O device fetches the processed data from the DMA and sends it to other entities.

\textbf{Discussion:} Both traditional MCU systems and MCUs equipped with a hardware security engine are characterized by frequent data movement. In the case of conventional MCU architecture, data is continuously shuttled between the CPU and SRAM, especially during processing operations such as encryption (step \Circled{\scriptsize{\textit{\textbf{4}}}}). Similarly, in an MCU system with a hardware security engine, data must also be frequently transferred between SRAM and the hardware engine (step \Circled{\scriptsize{\textit{\textbf{3}}}}-\Circled{\scriptsize{\textit{\textbf{4}}}}, \Circled{\scriptsize{\textit{\textbf{6}}}}-\Circled{\scriptsize{\textit{\textbf{7}}}}). Such extensive data movement not only introduces delay into the system but also results in unnecessary energy consumption.
In contrast, an MCU system with in-SRAM computing offers a more efficient solution. As all computation takes place within the SRAM, data merely needs to be stored in the SRAM initially and then transferred to the I/O device after processing. This approach effectively eliminates all unnecessary data movement, thereby reducing latency and power overheads.

\subsection{In-SRAM Computing}\label{sec:insram}

In-SRAM computing is a computational approach that executes processing tasks directly within the SRAM. It is an emerging technique that capitalizes on the intrinsic computational capabilities of SRAM to perform logic operations. By allowing for computational tasks to be performed within the memory where the data is stored, it considerably reduces the energy cost and latency associated with data movement between the processor and memory, which is common in traditional computing architectures.

\begin{figure}[!ht]
\centerline{\includegraphics[width=2.4in]{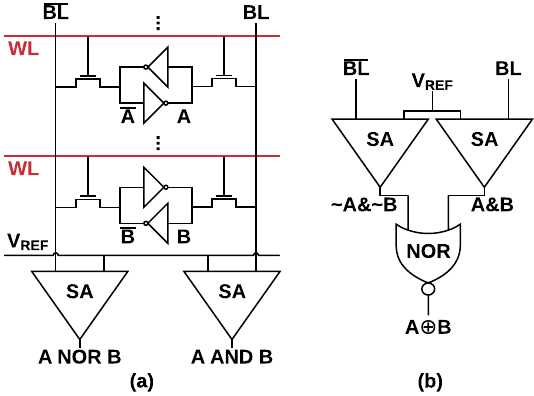}}
\caption{In-SRAM bitline operations \cite{zhangSealerInSRAMAES2022a}. 
}
\label{fig:bitline}
\end{figure}

In-SRAM processing is capable of bitline computing \cite{jeloka201628} by activating more than one row in the SRAM subarray. A diagram of the logical operations that can be performed in 6T SRAM is shown in Fig. \ref{fig:bitline}. AND and NOR operations are realized in SRAM with the help of several activated wordlines and sense amplifiers (SAs), as displayed in Fig. \ref{fig:bitline}(a). 


If all the cells in the active rows that are wired to the bitline $BL$ have the value `1', then the SA on the BL will be able to detect a voltage larger than $V_{ref}$, which is an element-wise AND operation.
Only if all the cells in the activated rows connected to the corresponding  $\overline{BL}$  contain `1', the SA on the $\overline{BL}$ senses a voltage greater than $V_{ref}$, which in turn requires that all the cells in the activated rows connected to the corresponding $BL$ contain `0'. This means the SA can do element-wise NOR operations. As can be seen in Fig. \ref{fig:bitline}(b), the XOR operation can be accomplished by combining the capabilities of the logical bit-wise AND and NOR operations.

Research has extended the capabilities of in-SRAM computing. For example, Cache Automaton \cite{subramaniyan2017cache} uses a sense-amplifier cycling mechanism to read out multiple bits in a single time slot, hence reducing input symbol match time. Compute Cache \cite{aga2017compute} increases the number of logical operations by modifying the SA architecture based on the NOR, AND, and XOR operations stated in \cite{jeloka201628}. 

Our approach, referred to as \textit{CryptoSRAM}, builds upon these techniques by utilizing the XOR capability along with 1-bit shifting, as detailed in Section \ref{sec:microarch}. By employing in-SRAM computing in MCU systems, we can significantly improve their throughput and energy efficiency, especially crucial in handling demanding IoT applications.


\section{System Integration Analysis}\label{sec:feasibility}

In-SRAM computing is a promising technique for boosting the performance and energy efficiency of systems. However, in non-Microcontroller Unit (MCU) systems, such as PCs and servers, enabling in-SRAM computing can lead to high overheads. This is primarily because in-SRAM computing requires operands to be orderly arranged within SRAM subarrays to endure data alignment according to computational needs, which is challenging to achieve efficiently in non-MCU systems.
Current approaches either neglect system integration issues \cite{zhangInhaleEnablingHighPerformance2022c} or require manual data layout \cite{zhangRecryptorReconfigurableCryptographic2018e}, presenting a gap in automated and efficient data arrangement.

\begin{figure}[!ht]
\centerline{\includegraphics[width=2.8in]{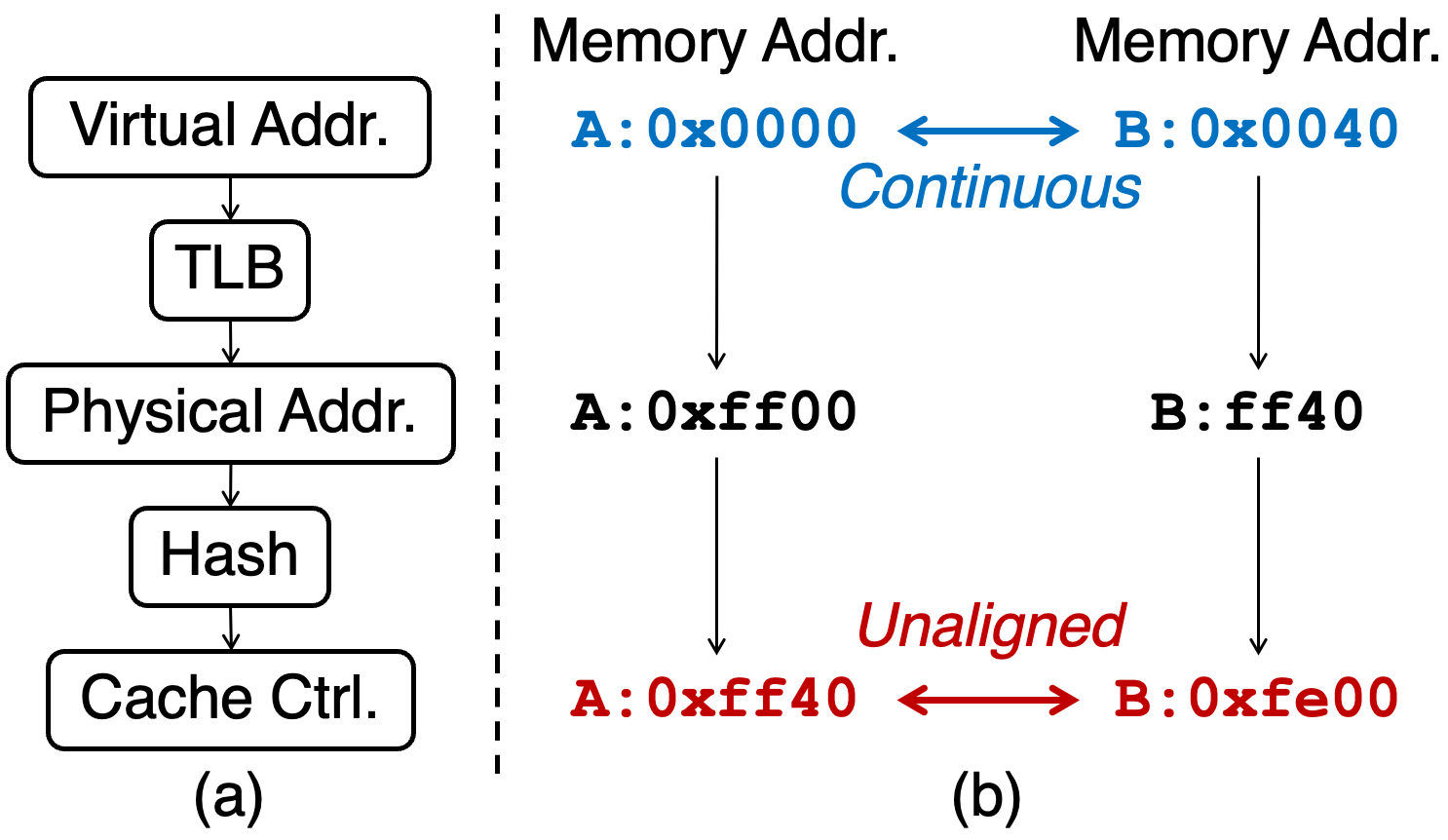}}
\caption{(a) Process of address translation in non-MCU systems. (b) Example of virtual address translation which makes the data layout uncontrollable in SRAM (i.e., cache in non-MCU systems).
}
\label{feasability}
\end{figure}

In most non-MCU systems, SRAM is primarily employed as a cache for the CPU. However, the prevalent use of technologies like virtual addressing, set-associative cache, and address hashing often makes it challenging for programs to dictate the precise data location within the SRAM.
Let's consider the deployment of virtual addressing which leads to an unpredictable data layout in the SRAM, as shown in Figure \ref{feasability}. Assume that a program intends to write two consecutive 2-byte data units (\bera{A} and \bera{B}) into the memory one by one. As depicted in Figure \ref{feasability}(a), the CPU first retrieves the virtual address of the data from the program. It then feeds this virtual address to the Translation Lookaside Buffer (TLB) to convert it into a physical address. This physical address is subsequently hashed to generate the final physical address that the cache controller uses to record the data. 
Despite \bera{A} and \bera{B} being located in successive memory locations from the programming view, post-TLB and hashing operations, their addresses get "randomly" mapped to different locations, as illustrated in Figure \ref{feasability}(b).
While virtual addressing simplifies programming by offering a unified and continuous memory space, it also relinquishes the control over the physical location of data in the SRAM (cache) and memory from the program's perspective. As a result, implementing in-SRAM computing becomes a challenge in non-MCU systems, unless there are significant modifications to the operating system or the memory allocator.

Conversely, in MCU systems, SRAM is used as memory. Furthermore, to enhance energy efficiency, MCU systems generally do not employ an operating system and virtual addressing or hashing. As a result, programs can specify the location of data in memory, i.e., SRAM. By using non-interleaved addressing \cite{12InterleavingOptions}, it becomes possible to place different bits of the same operand within the same subarray. Additionally, the use of DMA allows for internal data movement within SRAM, further increasing the flexibility of in-SRAM computing.
Therefore, in an MCU system, the integration of high parallelism and energy-efficient in-SRAM computing can be readily achieved with only bitline-computing-enabled SRAM subarrays and corresponding control logic. This inherent compatibility makes MCU systems a natural and low-overhead platform for enabling in-SRAM computing.

\section{Design of \textit{CryptoSRAM}}\label{implementation}
The cornerstone of our work lies in the novel implementation of an MCU system with inherent in-SRAM computing capabilities and its system integration (as detailed in this section) and algorithm co-designing (as detailed in section \ref{alg-co-design}).

\begin{figure*}[!ht]
\centerline{\includegraphics[width=6.9in]{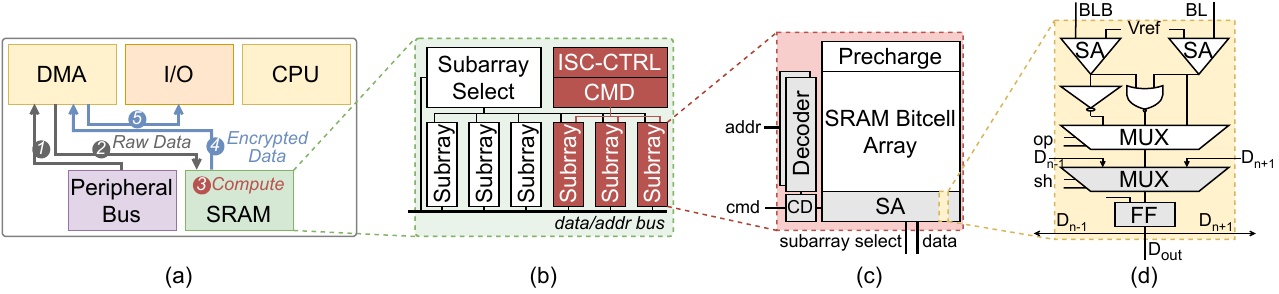}}
\caption{(a) Data path of sensor data in MCU with in-SRAM computing. Data movement between CPU and SRAM is avoided compared to conventional MCU systems.
(b) The structure of SRAM with computing capability. 
Portion of subarrays are modified to enable in-SRAM computing. In-SRAM control (ISC-CTRL) module and command (CMD) array are added to feed commands into ISC-enabled subarrays. 
(c) The structure of an ISC-enabled subarray with modified decoder (Decoder), sense amplifier (SA), and added command decoder (CD). 
(d) The structure of the modified SA supporting AND/OR/XOR/NOT logic and 1-bit shift operations. 
}
\label{microarch-all}
\end{figure*}

\subsection{MCU with In-SRAM Computing}\label{sec:microarch}

To efficiently tackle the challenges posed by conventional MCU architectures and hardware security engines, we propose and implement a unique MCU system that leverages in-SRAM computing. The pivotal change lies in the modification of the memory-centric SRAM of the MCU system to incorporate computational abilities.

As depicted in Figure \ref{microarch-all}(a), the data path of sensor data in our MCU system with in-SRAM computing has been substantially altered compared to traditional MCUs. In typical MCUs, sensor data must frequently move between the CPU and SRAM to execute operations such as encryption, leading to substantial overhead. In contrast, our proposed MCU system avoids such unnecessary data transfers between the SRAM and CPU.
In our system, raw data is first stored into the SRAM via DMA, as illustrated in steps \Circled{\scriptsize{\textit{\textbf{1}}}}-\Circled{\scriptsize{\textit{\textbf{2}}}} of Figure \ref{microarch-all}(a). Subsequently, within the SRAM, the raw data undergoes encryption through AES, with the authentication messages assembled concurrently (step \Circled{\scriptsize{\textit{\textbf{3}}}}). This step leverages the computational capabilities of our modified SRAM, avoiding the need for data transfer to the CPU. Finally, the encrypted data, coupled with authentication messages, is transmitted to I/O devices via DMA (steps \Circled{\scriptsize{\textit{\textbf{4}}}}-\Circled{\scriptsize{\textit{\textbf{5}}}}).
A closer look at the structure of our computing-capable SRAM is provided in Figure \ref{microarch-all}(b). Here, we observe the modifications to a portion of the SRAM subarrays to enable in-SRAM computing. We add an In-SRAM Control (ISC-CTRL) module and a command (CMD) array to guide commands into the ISC-enabled subarrays, which are discussed in Section \ref{sec:control}.

To facilitate in-SRAM computing in the MCU system, we make essential microarchitectural changes by performing several low-overhead modifications to the standard SRAM subarray.
Compared to a standard SRAM subarray, an In-SRAM Computing (ISC)-enabled SRAM subarray introduces an additional row decoder and a command decoder (CD), as shown in Figure \ref{microarch-all}(c). Additionally, we have modified the sense amplifier, as depicted in Figure \ref{microarch-all}(d).

\textbf{Decoder:} The additional row decoder is utilized to implement the simultaneous activation of two wordlines, as mentioned in Section \ref{sec:insram}. 
For example, we want to perform a logic operation in ISC-enabled subarray with \textit{act\_row} and \textit{logic\_op} commands (details will be discussed later).
When logic operations are needed, \textit{act\_row} sends the row index of the first wordline to be activated to the first row decoder, while \textit{logic\_op} dispatches the row index of the second wordline to the second row decoder.  After both row decoders have received row indexes, they simultaneously activate the two wordlines.

\textbf{Command decoder:} Another crucial addition is a command decoder (CD) used to decode received commands into control signals for row decoders and sense amplifiers. For instance, the \textit{act\_row} and \textit{logic\_op} commands, with their address \textit{src1} and \textit{src2} (Table \ref{cmd-table}), should be decoded by CD, leading to the activation of two different wordlines. Additionally, the other bits in the commands are decoded into control signals to govern various select signals in the sense amplifier. The 2-bit \textit{op}, which controls the type of logic operation, is derived by decoding the lower 4 bits of the \textit{logic\_op} command.

\textbf{Sense amplifier: }The sense amplifier (SA) is modified to support basic logical computations and 1-bit shifts, as shown in Figure \ref{microarch-all}(d). A multiplexer (MUX) and a flip-flop (FF) have been added to enable 1-bit shift operations. This configuration maintains the integrity of the 1-bit shift operations.

\textbf{Control Commands: } In total, we use six commands: \textit{rd\_row}, \textit{wr\_row}, \textit{shift}, \textit{act\_row}, \textit{logic\_op}, and \textit{ext\_bit}, as listed in Table \ref{cmd-table}. The \textit{rd\_row} command is used to read the data from a specific row to the SA or the data bus. The \textit{wr\_row} command writes the data from the SA or data bus to a particular row. After executing the \textit{rd\_row} command, the \textit{shift} command performs a 1-bit shift of the data in the SA either left or right. The middle 8 bits specify how many times the 1-bit shift operation should be executed, making the command more compact. The \textit{act\_row} and \textit{logic\_op} commands are usually used together. The \textit{act\_row} command first informs one row address to the first decoder. Then, \textit{logic\_op} informs another row address to the second decoder, achieving the simultaneous activation of two wordlines. Also, \textit{logic\_op} conveys the type of logic operation to the SA by using the CD to decode the rightmost four bits of \textit{logic\_op}.
In the \textit{ext\_bit} command, the index bits represent the column index. By using a fixed row for bit extension (e.g., the last row), the command with the column index can select the corresponding bitline. The 3-bit in \textit{ext\_bit} control field specifies the computing block width (16/32/64/128/256/512). 
These six commands hence facilitate precise control of bitline computing. The addition of one row decoder, a CD, and modifications in the SA have enabled in-SRAM computing in MCU systems.

\begin{table}[ht]
	\centering
	\caption{The formats of the ISC control commands. Each command has 4 bits of opcode, 8 bits of address (shifted number in \textit{shift} command), and 4 bits of variant indicator.}
	\scalebox{1.0}{
	\begin{tabular}{>{\centering\arraybackslash}p{0.4in}>{\centering\arraybackslash}p{0.4in}>{\centering\arraybackslash}p{0.4in}>{\centering\arraybackslash}p{0.4in}>{\centering\arraybackslash}p{0.9in}}
		\toprule
           CMD & Opcode (4b) & Index (8b) & Option (4b) & Function\\
		\midrule
            rd\_row & 0001 & src & 1000 & read row\\
            wr\_row & 0010 & dst & x000 & write row\\
            shift & 0011 & num & 1xx0 & left/right shift\\
            act\_row & 1011 & src1 & 0001 & activate wordline\\
            logic\_op & 1001 & src2 & 0xx0 & logic compute\\
            ext\_bit & 1111 & col & xxx0 & extend bit\\
		\bottomrule
	\end{tabular}}
	\label{cmd-table}
\end{table}

\subsection{Control Scheme of \textit{CryptoSRAM}}\label{sec:control}

This subsection introduces on the control scheme that enables task offloading from the CPU to the SRAM in the context of the \textit{CryptoSRAM} system. We achieve this by incorporating an In-SRAM Computing Control Module (ISC-CTRL) and a Command Array (CMD) in the SRAM, as depicted in Figure \ref{microarch-all}(b). These additions become operational once the data is moved from the sensor to the ISC-enabled subarray by DMA using the corresponding physical addresses.

\begin{figure}[!ht]
\centerline{\includegraphics[width=2.8in]{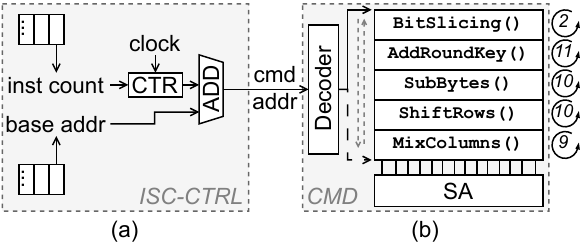}}
\caption{The control mechanism to generate instructions for AES-128 loops of 2$\times$ \bera{BitSlicing}, 11$\times$ \bera{AddRoundKey}, 10$\times$ \bera{SubBytes}, 10$\times$ \bera{ShiftRows}, 9$\times$ \bera{MixColumns}. 
}
\label{ctrl}
\end{figure}

Once the DMA completes transferring data from the sensor to the subarray, ISC-CTRL initiates the process by controlling CMD to read and send the appropriate command to the CD of the subarray. It's worth noting that due to the use of non-interleaved addressing, data can be written into another subarray while one subarray is performing computations. 

The operation of ISC-CTRL is detailed in Figure \ref{ctrl}. The control module consists of a buffer that stores the base addresses of command sets corresponding to various functions, and another buffer that holds the number of commands each function includes. These command sets are stored in the CMD array, which can be accessed by ISC-CTRL to facilitate the execution of tasks in the ISC-enabled subarray.

The control mechanism works by first reading the base address and the instruction count of a function's command set. It uses these values to calculate the address of the command required in the command array for that specific cycle. The instruction count is used to set the maximum number of the counter (CTR), facilitating the efficient repetition of the function in a low-overhead manner.

\section{Algorithm-architecture Co-designing} \label{alg-co-design}

In this section, we discuss the co-design of algorithm and architecture for in-SRAM computing. This co-design is crucial for optimizing the system performance in terms of processing speed, power consumption, and overall system efficiency. A critical factor in this optimization process is the layout of data within the SRAM array. As bitline computing requires operands to share the same bitline, meaning they must be in the same column, frequent alterations to the data layout during computation can lead to significant overhead from data movement, primarily shift operations. Therefore, an optimal data layout can greatly reduce the data movement overhead in in-SRAM computing.

Here, we will be presenting the mapping of Advanced Encryption Standard (AES) and Secure Hash Algorithm 3 (SHA3) onto in-SRAM computing. The choice of these algorithms is motivated by their widespread use and adaptability. AES, as the most widely used block encryption algorithm, plays a critical role in ensuring secure data transmission. SHA3, on the other hand, offers versatility with its robust security margin and adaptable output lengths, thereby providing suitability to varied security levels and device capabilities—an essential feature in diverse IoT environments.

\subsection{Computing Block}

\begin{figure}[!ht]
\centerline{\includegraphics[width=2.8in]{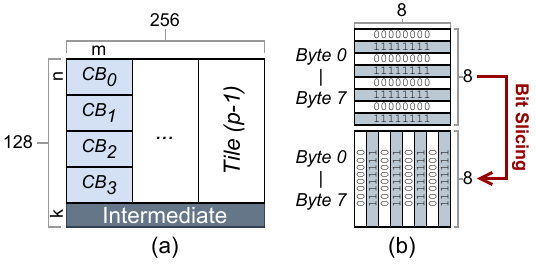}}
\caption{(a) Data organization in an ISC-enabled subarray.
All the variables ($k,m,n$) are configurable to accommodate different algorithms in different settings and modes. 
(b) An AES example of data mapping in a CB with bit slicing technique. 
}
\label{aes-mapping}
\end{figure}

To support various algorithms with different data widths, we employ Computing Blocks (CBs) as fundamental computational units, as illustrated in Fig. \ref{aes-mapping}(a). Each CB consists of \textit{n} rows and \textit{m} columns of SRAM cells, with dimensions varying based on algorithm requirements. 
Our design uses \textit{k} rows for intermediate variable storage, with \textit{k} flexibly adjusted based on application needs. Furthermore, CBs that share the same set of bitlines form a tile, resulting in $p=\left \lfloor 256/m\right \rfloor$ tiles for a 256-column ISC-enabled subarray. CBs that share the same wordline in different tiles can execute parallel operations.

\subsection{Advanced Encryption Standard}

\begin{figure}[!ht]
\centerline{\includegraphics[width=3.2in]{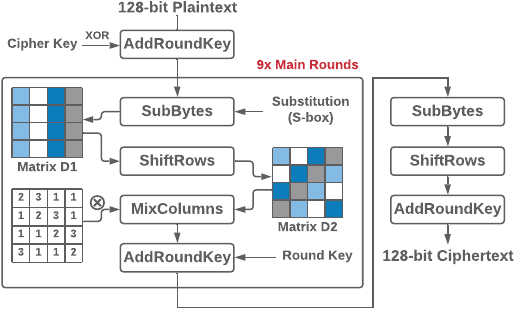}}
\caption{The steps of AES for a 128-bit plaintext. This figure illustrates AES in Electronic Codebook (ECB) mode - the most basic form of AES encryption. Other complex modes like Cipher Block Chaining (CBC), Counter with CBC-MAC (CCM), and Galois/Counter Mode (GCM) also exist, enhancing security by building upon the ECB structure \cite{zhangSealerInSRAMAES2022a}. 
}
\label{AES_flow}
\end{figure}

The Advanced Encryption Standard (AES) in cryptography, also known as Rijndael cryptography, is a block encryption standard.
The AES encryption process operates on a 4×4 matrix of bytes
whose initial value is a plaintext block (the element size in the matrix is one byte). 
Each AES encryption round (except the last one) consists of four steps where the output of each stage is used as an input for the next stage and described as follows.  
(1) \bera{AddRoundKey}: each byte in the matrix is XORed with the round key. Each round key is generated by the key generation scheme from a given cipher key. 
(2) \bera{SubBytes}: each byte is substituted with its corresponding byte in the S-box block using a non-linear substitution function commonly implemented with lookup tables (LUTs). 
(3) \bera{ShiftRows}: a round-robin shift is performed for each row in the matrix. The first row is unchanged while each element of the second row are shifted to the left by 1 byte. Then for the third and fourth row, elements are shifted to the left by 2 and 3 bytes, respectively (see matrix D1 to D2 transformation in Figure \ref{AES_flow}). 
(4) \bera{Mixcolumns}: this step uses a linear transformation to fully mix the four bytes of each column. In AES, the \bera{MixColumns} stage is omitted from the last encryption loop and replaced with another \bera{AddRoundKey}.

The overall process of AES for a 128-bit plaintext is shown in Figure \ref{AES_flow}. First, the plaintext is XORed with the original cipher key for initialization. Then after 9 main rounds, each of which consists of \bera{AddRoundKey}, \bera{SubBytes}, \bera{ShiftRows}, and \bera{MixColumns}, and the final round without \bera{MixColumns}, the 128-bit ciphertext is generated.

\textbf{Mapping: }
We explain
our method of mapping the AES algorithm onto in-SRAM computing. As shown in Figure \ref{aes-mapping}(b), the data bytes are stored in the SRAM array row by row, with each row holding a byte, managed by the DMA. We then employ bit slicing \cite{benadjilaImplementingLightweightBlock2013}, transposing the data block so that bits in the same position from different bytes occupy the same row. 
Bit slicing is a technique employed in computer systems design that involves breaking down computations into simple bitwise operations that can be performed in parallel. This is achieved by treating each bit of data individually, rather than as part of a larger unit.

After bit slicing operation, we execute the various rounds of the AES algorithm.
During the \bera{AddRoundKey} phase, the data block undergoes a XOR operation with the corresponding bit of the pre-stored key in the SRAM, utilizing bitline computing. In the \bera{SubBytes} phase, we avert the necessity for high-overhead lookup table-based computation due to our use of bit slicing. Instead, we employ a computation method based on combinational logic \cite{boyarNewLogicMinimization}. The \bera{ShiftRows} phase follows, where, thanks to bit slicing, byte-level shifts are achieved by executing a maximum of 3-bit shifts left or right on each row.
Finally, in the \bera{MixColumn} phase, as we position different bits on different rows, bit-level operations are performed with varying computations on different bits (for example, executing 4 XORs on the first bit of byte \textit{A} and 6 XORs on the second bit). This prevents the redundancy of byte-level computations.
In summary, we employ bit slicing to map AES onto in-SRAM computing, significantly reducing data movement overhead and improving parallelism to the maximum extent.

In a typical AES algorithm implementation that leverages the bit slicing technique, four 16-byte data blocks are interleaved to form an 8$\times$64 array for computation \cite{PQClean2023}. This is because a 64-bit data width (column count) can efficiently utilize the 32-bit or 64-bit registers in the CPU, thus enhancing performance.
However, this computation method does not take into account the fact that in-SRAM computing is adept at wide-width logical operations but not lengthy shift operations. To adapt to the characteristics of in-SRAM computing, we abandon the preprocessing stage of interleaving four data blocks. Instead, we directly apply bit slicing operations to non-interleaved data blocks. This change eliminates the overhead associated with lengthy shift operations and transforms them into shorter shift operations, improving overall performance.
As per our experiments, the in-SRAM AES implementation without CPU-optimized interleaving outperforms the in-SRAM AES implementation with CPU-optimized interleaving by approximately 46\% in terms of latency.

In the implementation of the AES algorithm, we designate the parameters of the computing block as $n=8$, $m=16$, and $k=100$. The selection of these parameters is designed to facilitate the efficient storage of precomputed keys and intermediate variables.
With this configuration, the value of $p$, which represents the number of tiles in a subarray, is 16. This implies that in a 128$\times$256 ISC-enabled subarray, up to 16 data blocks can undergo AES computation simultaneously.

\subsection{Secure Hash Algorithm 3}

\begin{figure}[!ht]
\centerline{\includegraphics[width=2.8in]{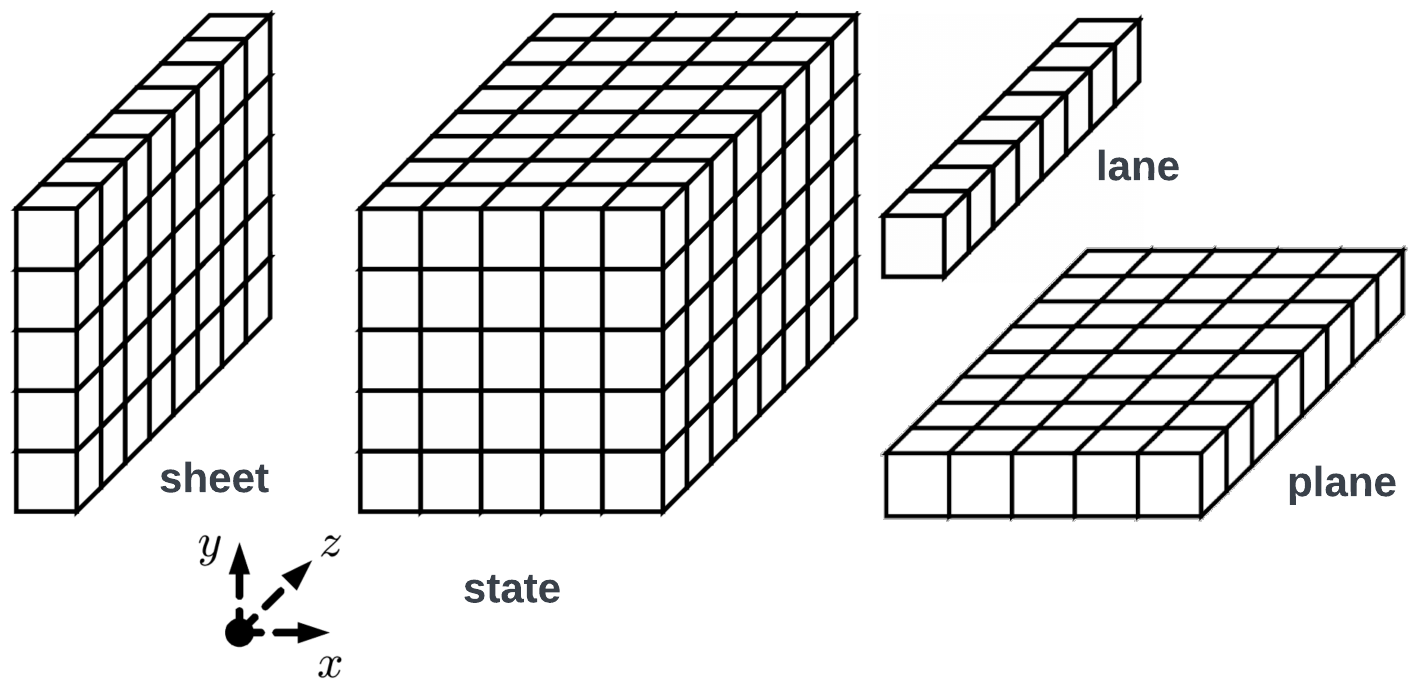}}
\caption{Naming conventions for parts of Keccak-$f$ \textit{State} \cite{dworkin_sha-3_2015}.
The Keccak-$f$[1600] function, fundamental to SHA-3, utilizes a 1600-bit state configured as a 3D array (5$\times$5$\times$64). Each element in this array, referred to as a \textit{lane}, is a 64-bit word. The algorithm applies a sequence of five core steps—$\theta$, $\rho$, $\pi$, $\chi$, and $\iota$—to this multi-\textit{lane} structure, ensuring a robust level of security.
}
\label{namingstate}
\end{figure}

The construction and verification of digital signatures, key derivation, and the generation of pseudo-random bits are the primary functions of the hash algorithm, which is an essential component of the information security domain.
SHA-3 is the third generation of standard hash functions, based on the implementation of the Keccak algorithm. The SHA-3 algorithm, unlike its predecessors, is a permutation-based cryptographic function.
SHA-3 hashing heavily relies on the structure of the Keccak sponge function.
The sponge structure is capable of data transformation, i.e., transforming arbitrary-length inputs into arbitrary-length outputs.
The fundamental function in Keccak is a permutation selected from the set of seven Keccak-$f$ permutations, abbreviated Keccak-$f[b]$, where $b\in\{25,50,100,200,400,800,1600\}$ is the width of the permutation.
In the sponge structure, the width of the permutation corresponds to the width of the \textit{State}.
The \textit{State} is organized as a grid of five-by-five \textit{lanes} whose length is $w\in\{1,2,4,8,16,32,64\}$ and $b=25w$. 
The naming conventions for parts of the Keccak-$f$ \textit{State} used throughout the paper are shown in Fig. \ref{namingstate}.

Keccak has five stages of computing ($\theta,\rho,\pi,\chi,\iota$) in one round. Let's take one round of Keccak-$f[1600]$ as an example. Suppose we have a 1600-bit input \textit{State}, which according to the naming conventions (Fig. \ref{namingstate}), we can refer to as $A[x,y,z]$, where $x,y\in{0...4},z\in{0...63}$. The computations in five stages are as follows \cite{dworkin_sha-3_2015}:
\begin{enumerate}
    \item[$\theta$:] $C[x] = parity(A[x,0...4]), \hfill \forall x \in \{0...4\}$\\
                     $CT[x] = C[x] <<< 1, \hfill \forall x \in \{0...4\}$\\
                     $FT[x] = C[x-1] \oplus CT[x+1], \hfill \forall x \in \{0...4\}$
                     $A[x,y] = A[x,y] \oplus FT[x], \hfill \forall (x,y) \in \{0...4,0...4\}$
    \item[$\rho$:] $A[x,y] = A[x,y] <<< r(x,y), \hfill \forall (x,y) \in \{0...4,0...4\}$
    \item[$\pi$:]  $B[y,2x+3y]=A[x,y], \hfill \forall (x,y) \in \{0...4,0...4\}$
    \item[$\chi$:] $A[x,y]=B[x,y]\oplus (\neg B[x+1,y] \wedge B[x+2,y]),$\\
                    \hspace*{0pt}\hfill $\forall (x,y) \in \{0...4,0...4\}$
    \item[$\iota$:] $A[0,0]=A[0,0] \oplus RC[i]$,
\end{enumerate}
where the constant $r(x,y)$ in the $\pi$ stage is the rotation offset depending on the location of the \textit{lane}, and $RC[i]$ in the $\iota$ stage is the round constant which is different for each of 24 rounds in Keccak-$f[1600]$. For different $w$, the number of rounds is equal to $12+2log_2w$. The output of each round will be XORed with the next piece of the message (1088 bits in Keccak-$f[1600]$) as the input of next round of hashing \cite{dworkin_sha-3_2015}.

\textbf{Mapping: }
We explain our method of mapping the SHA-3 algorithm onto in-SRAM computing. For keccak-\textit{f}[1600], we place the 25 \textit{lanes} of 64-bits each in the same column but different rows of the 25 rows \cite{zhangInhaleEnablingHighPerformance2022c}. During the $\theta$ phase, XOR operations are performed between \textit{lanes} without the need for shifts. Rotate shifts are achieved by combining two shift operations and an OR operation (for example, left rotate shift 16 bits, which involves logically shifting left by 16 bits, storing the results in an intermediary row, then performing a logical right shift of 48 bits on the original \textit{lane}, and finally executing an OR operation on the two shifted \textit{lanes}). This method of rotate shift is also applicable to the $\rho$ phase.
In the $\pi$ stage of SHA3, a shift operation involving the entire 64-bit data \textit{lane} and other \textit{lanes} is necessary. With the \textit{lane-per-row} data layout, we are able to use the row index to select the corresponding data \textit{lane}, thereby eliminating the need for \textit{inter-lane} shifts. This results in more efficient shift operations by avoiding time-consuming shuffling of data across different \textit{lanes}.
Our analysis indicates that approximately 90\% of shift overhead can be eliminated using this design.
For the $\chi$ and $\iota$ phases, we can carry out the computations using \textit{lane}-level XOR/OR/NOT operations.
In summary, by employing the \textit{lane-per-row} technique to map SHA-3 onto in-SRAM computing, we eliminate the need for \textit{inter-lane} shift operations and significantly enhance performance.


In order to efficiently accommodate the SHA3 algorithm, we configure the computing blocks (CBs) with $n = 25$ and $m = 64$, and set $k = 6$ to store intermediate variables. These settings align with the requirements of SHA3 for computation and storage.
With these parameters, there are $p = 4$ tiles within a 128$\times$256 subarray, meaning that up to four SHA3 computations can be conducted simultaneously in one subarray. This significantly speeds up the processing time, and therefore enhances the overall system throughput.
Moreover, the unused portions of the subarray not occupied by the data blocks can be effectively used to store additional data for future SHA3 calculations. This utilization approach makes the most of the available memory resources and helps to reduce the storage overhead, further improving the system's efficiency and cost-effectiveness.

\section{Evaluation}\label{evaluation}

\begin{table*}[!ht]
\centering
\caption{Throughput over Different AES Modes}
\label{tab:aes-throughput}
\scalebox{1.0}{
\begin{tabular}{lccc|ccc|ccc|ccc}
\toprule
& \multicolumn{6}{c}{\textbf{AES-Encrypt}} & \multicolumn{6}{c}{\textbf{AES-Decrypt}} \\
\textbf{Throughput (MB/s)} & \multicolumn{3}{c}{\textbf{AES-128}} & \multicolumn{3}{c}{\textbf{AES-256}} & \multicolumn{3}{c}{\textbf{AES-128}} & \multicolumn{3}{c}{\textbf{AES-256}} \\
& CBC & CCM & GCM & CBC & CCM & GCM & CBC & CCM & GCM & CBC & CCM & GCM \\
\midrule
CPU (Software) & 1.448 & 0.86 & 0.876 & 1.145 & 0.654 & 0.756 & 1.32 & 0.858 & 0.878 & 1.061 & 0.653 & 0.758 \\
ASIC (Hardware) & 17.543 & 9.661 & 15.847 & 13.55 & 7.507 & 12.437 & 17.361 & 9.718 & 15.6 & 13.404 & 7.535 & 12.269 \\
\textit{CryptoSRAM} (25\%) & 28.337 & 14.169 & 10.999 & 21.406 & 10.703 & 9.771 & 24.208 & 12.104 & 10.316 & 18.086 & 9.043 & 9.016 \\
\textit{CryptoSRAM} (50\%) & 56.674 & 28.337 & 21.998 & 42.813 & 21.406 & 19.542 & 48.416 & 24.208 & 20.632 & 36.172 & 18.086 & 18.031 \\
\textit{CryptoSRAM} (100\%) & 113.348 & 56.674 & 43.996 & 85.625 & 42.813 & 39.084 & 96.832 & 48.416 & 41.264 & 72.344 & 36.172 & 36.062 \\
\bottomrule
\end{tabular}}
\end{table*}


\subsection{Evaluation Methodology}\label{sec:methodology}

All results are evaluated on the STM32L562 \cite{STM32L5x2ArmCortex} microcontroller, which features an Arm Cortex-M33 32-bit RISC core, up to 512 KB of Flash memory, and 256 KB of SRAM. The STM32L562 includes an embedded hardware security accelerator, making it well-suited for secure firmware due to its efficient cryptographic operations.
Benchmarks for CPU and ASIC implementations are generated with CyclonCRYPTO 2.1.6 and compiled using Clang with optimization level 3 \cite{STM32L5CryptoBenchmark}. Power specifications for MCU modes follow \cite{STM32L5SystemPowerControlPWR}. The STM32L562 is operated at 1.8V VDD and 25 \textdegree C, with evaluations conducted across three operational modes: (I) RUN mode (Range0) at 110 MHz and 11.21 mA with all peripherals enabled, (II) RUN mode (Range2) at 26 MHz and 1.87 mA with limited peripherals enabled, and (III) Low-power SLEEP mode at 2 MHz and 230 $\mu$A, where the CPU clock is off and Flash can be completely unpowered for energy savings. In all modes, the Random Number Generator is assumed active to generate initialization vectors (IVs) for ensuring unique ciphertexts.

For SRAM-based in-memory secure computation (\textit{CryptoSRAM}), a custom cycle-accurate simulator was developed, integrating power, area, and timing metrics derived from synthesis results using Synopsys Design Compiler and Cadence Innovus (45 nm node). SRAM circuits were generated with Verilog and CNC layouts using OpenRAM and PyMTL3. SPICE simulations validated the timing and power of SRAM bitline operations. Only in-SRAM components were simulated, with CPU and ASIC data sourced from hardware specifications. The simulator models \textit{CryptoSRAM} performance using the same frequency and power metrics as the STM32L562. To account for varying ratios of ISC-enabled SRAM subarrays, we conservatively assume equal power consumption across all subarrays, though power gating can reduce inactive subarray consumption.
ASIC designs are evaluated optimistically by excluding power overheads introduced by additional hardware components. This ensures a fair comparison, highlighting the advantages of the proposed in-SRAM computing scheme.

\subsection{Throughput}
Table \ref{tab:aes-throughput} presents the encryption and decryption throughput comparison among the CPU, ASIC, and \textit{CryptoSRAM} for different AES modes, respectively. All the throughput analysis is running on the high-performance MCU system mode (RUN (Range0)) with 110 MHz frequency.
\textit{CryptoSRAM} generally outperforms CPU and ASIC implementations across various modes of AES, including AES-128-CBC, AES-128-CCM, AES-128-GCM, AES-256-CBC, AES-256-CCM, and AES-256-GCM.
For AES-128-CBC and AES-128-CCM modes, \textit{CryptoSRAM} (even at 25\% capacity) offers a throughput that is almost 1.5$\times$ greater than that of ASIC (the fastest of the traditional implementations) and substantially surpasses the throughput of the CPU by a factor of nearly 19 and 16 respectively. Similarly, the AES-256-CBC and AES-256-CCM modes demonstrate comparable improvements.
\textit{CryptoSRAM} boasts a higher number of vector processing units (inherent from in-SRAM computing, Section \ref{sec:insram}), compared to ASIC implementations. This enables \textit{CryptoSRAM} to process data in parallel at a much larger scale, resulting in higher throughput performance.

The exception to this pattern is the AES-128-GCM and AES-256-GCM modes, where the 25\% capacity \textit{CryptoSRAM} implementation underperforms relative to the ASIC version by roughly 31\% and 22\%, respectively. This is due to the fact that the conditional statement in AES-GCM is challenging to efficiently execute with in-SRAM computing. Nonetheless, as the proportion of ISC-enabled SRAM rises to 50\% and 100\%, the \textit{CryptoSRAM} outperform both the CPU and ASIC implementations, including AES-GCM modes. Specifically, with 50\% ISC-enabled SRAM, the \textit{CryptoSRAM} configuration delivers a performance improvement of 36\% and 54\% respectively over ASIC version for both AES-128-GCM and AES-256-GCM.

\begin{table}[tp]
\centering

\caption{Throughput over Different SHA3 Variants}
\label{tab:sha3-throughput}
\scalebox{0.95}{
\begin{tabular}{>{\centering\arraybackslash}p{1.1in}>{\centering\arraybackslash}p{0.4in}>{\centering\arraybackslash}p{0.4in}>{\centering\arraybackslash}p{0.4in}>{\centering\arraybackslash}p{0.4in}}
\toprule
\textbf{Throughput (MB/s)} & \textbf{SHA3-224} & \textbf{SHA3-256} & \textbf{SHA3-384} & \textbf{SHA3-512} \\
\midrule
CPU (Software) & 0.893 & 0.844 & 0.648 & 0.45 \\
\textit{CryptoSRAM} (25\%) & 15.098 & 14.260 & 10.904 & 7.549 \\
\textit{CryptoSRAM} (50\%) & 30.197 & 28.519 & 21.809 & 15.098 \\
\textit{CryptoSRAM} (100\%) & 60.393 & 57.038 & 43.617 & 30.197 \\
\bottomrule
\end{tabular}}
\end{table}

Table \ref{tab:sha3-throughput} presents the throughput comparison between the CPU and \textit{CryptoSRAM} for SHA3 with different configurations. Due to the absence of an ASIC design for SHA3 in existing systems \cite{STM32L5CryptoBenchmark}, the comparison was made solely with the CPU's software implementation.
For all configurations (SHA3-224, SHA3-256, SHA3-384, and SHA3-512), \textit{CryptoSRAM} surpasses the CPU's performance by a significant margin. Even at 25\% capacity, \textit{CryptoSRAM} outperforms the CPU by factors of around 17 for SHA3 with different configurations. The throughput of \textit{CryptoSRAM} increases linearly with the portion of ISC-enabled SRAM, doubling when capacity is increased from 25\% to 50\% and doubling again from 50\% to 100\%. This indicates a highly scalable system performance that can be adjusted according to the available computing resources in the SRAM.

In Table \ref{hmac}, we present the throughput performance of \textit{CryptoSRAM} for SHA3-based HMAC (Hash-based Message Authentication Code). For the evaluation, we use keys and messages of the same length as the corresponding SHA3 variants. The performance of \textit{CryptoSRAM} in SHA3-based HMAC closely mirrors its performance in SHA3. The key difference lies in the additional XOR operation performed on the key and the concatenation of keys and messages in HMAC. However, due to the lane-per-row data layout that we employ, the overhead incurred by these additional operations is negligible.

\begin{table} [tp]
    \centering
    
    \caption{HMAC throughput for SHA3 variants. Lengths of key and message are the same as the SHA3 block sizes.}
    \scalebox{1}{
    \begin{tabular}{>{\centering\arraybackslash}p{1.1in}>{\centering\arraybackslash}p{0.4in}>{\centering\arraybackslash}p{0.4in}>{\centering\arraybackslash}p{0.4in}>{\centering\arraybackslash}p{0.4in}}
        \toprule
        \textbf{Throughput (MB/s)} & \textbf{SHA3-224} & \textbf{SHA3-256} & \textbf{SHA3-384} & \textbf{SHA3-512} \\
        \midrule
        \textit{CryptoSRAM} (25\%) & 4.034 & 3.810 & 2.914 & 2.017 \\
        \textit{CryptoSRAM} (50\%) & 8.069 & 7.621 & 5.828 & 4.034 \\
        \textit{CryptoSRAM} (100\%) & 16.138 & 15.241 & 11.655 & 8.069 \\
        \bottomrule
    \end{tabular}}
    \label{hmac}
\end{table}

\subsection{Energy Efficiency}



Table \ref{aes128cbc} illustrates a comparative evaluation of the energy efficiency, defined as throughput per unit power (GB/s/W), of our proposed \textit{CryptoSRAM} design versus CPU and ASIC hardware under various operating conditions. Energy efficiency, represented by the metric throughput per unit power, is a critical indicator of a system's performance as it demonstrates how much work (throughput) the system can perform for each unit of power consumed.

\begin{table}[tp]
    \centering
    \caption{Energy Efficiency of AES over Power Modes}
    \scalebox{1.0}{
    \begin{tabular}{lccc}
        \toprule
        \textbf{Tput./Power (GB/s/W)} & \textbf{110 MHz} & \textbf{26 MHz} & \textbf{2 MHz} \\
        \midrule
        CPU (Software) & 0.0718 & 0.1017 & 0.0000 \\
        ASIC (Hardware) & 0.8694 & 1.2319 & 0.7704 \\
        \textit{CryptoSRAM} (25\%) & 1.3396 & 1.8982 & 1.1872 \\
        \textit{CryptoSRAM} (50\%) & 2.6793 & 3.7963 & 2.3743 \\
        \textit{CryptoSRAM} (100\%) & 5.3586 & 7.5927 & 4.7486 \\
        \bottomrule
    \end{tabular}}
    \label{aes128cbc}
\end{table}

\begin{table}[tp]
    \centering
    \caption{Energy Efficiency of SHA3 over Power Modes}
    \scalebox{1.0}{
    \begin{tabular}{lccc}
        \toprule
        \textbf{Tput./Power (GB/s/W)} & \textbf{110 MHz} & \textbf{26 MHz} & \textbf{2 MHz} \\
        \midrule
        CPU (Software) & 0.0418 & 0.0593 & 0.0000 \\
        \textit{CryptoSRAM} (25\%) & 0.7067 & 1.0013 & 0.6262 \\
        \textit{CryptoSRAM} (50\%) & 1.4134 & 2.0026 & 1.2525 \\
        \textit{CryptoSRAM} (100\%) & 2.8267 & 4.0053 & 2.5050 \\
        \bottomrule
    \end{tabular}}
    \label{sha3-256}
\end{table}

In the context of AES-128-CBC running at 110 MHz, \textit{CryptoSRAM} (25\%) shows a throughput/power of about 1.34 GB/s/W, which is around 55\% higher than the 0.87 GB/s/W shown by the ASIC hardware. As the ISC-enabled SRAM proportion increases to 50\% and 100\%, the throughput/power leaps to approximately 2.68 GB/s/W and 5.36 GB/s/W respectively. This demonstrates that \textit{CryptoSRAM} is utilizing in-SRAM computing effectively to enhance energy efficiency.
At a lower frequency of 26 MHz, the energy efficiency of \textit{CryptoSRAM} (25\%) continues to outperform the ASIC design, achieving 1.90 GB/s/W compared to ASIC's 1.23 GB/s/W. This performance gap further widens as we enable more SRAM with ISC capabilities.
Interestingly, at 2 MHz, which represents a low-power sleep mode, the CPU is unable to compute, whereas the \textit{CryptoSRAM} (25\%) still maintains a throughput/power of 1.19 GB/s/W. 
Similar trends are observed for SHA3-256 in Table \ref{sha3-256}. At 110 MHz, \textit{CryptoSRAM} (25\%) shows a throughput/power of about 0.71 GB/s/W, which is significantly higher (around 16 times) than the 0.042 GB/s/W exhibited by CPU software. The difference further accentuates as the proportion of ISC-enabled SRAM increases to 50\% and 100\%.
In the low-power sleep mode, \textit{CryptoSRAM} maintains a strong performance with a throughput/power of 0.63 GB/s/W at 25\% ISC-enabled SRAM, whereas the CPU is unable to perform any computation.

\subsection{Control Overhead Analysis}
Table \ref{ctrl-overhead} provides a detailed breakdown of the control overhead for the AES-128, GHASH, and SHA3 functions. The table includes the number of instructions (\#Inst.), capacity requirements in kilobytes (KB), and the number of iterations (\#Iter.) for each function. For AES-128, the functions \bera{BitSlicing()}, \bera{AddRoundKey()}, \bera{SubBytes()}, \bera{ShiftRows()}, and \bera{MixColumns()} are included. Their capacity requirements range from 0.05 KB to 0.91 KB, while the storage requirement of \bera{ShiftRows()} is the highest. The functions of AES-128 also require a varying number of iterations, with \bera{AddRoundKey()} and \bera{ShiftRows()} having the most (11 and 10 respectively).
The GHASH function includes three different operations, \bera{ByteArrange()}, \bera{ByteAligning()}, and \bera{GaloisMult()}. The \bera{GaloisMult()} function stands out due to its significantly higher iteration count (1024), despite having the smallest capacity requirement (0.03 KB) among all the functions listed.
Lastly, SHA3 includes only the \bera{StatePermute()} function which requires 1.27 KB and needs to be iterated 24 times.
In total, the listed functions consist of 2233 instructions, requiring a total capacity of 4.47 KB. Therefore, for successful operation of these cryptographic functions, it is necessary to ensure that the capacity of the command array exceeds this requirement.

\begin{table} [!ht]
	\centering
	\caption{Control overhead of AES-128, GHASH, SHA3.}
	\scalebox{0.9}{
	\begin{tabular}{>{\centering\arraybackslash}p{0.4in}>{\centering\arraybackslash}p{0.9in}>{\centering\arraybackslash}p{0.3in}>{\centering\arraybackslash}p{1.0in}>{\centering\arraybackslash}p{0.3in}}
		\toprule
            & \textbf{Function} & \textbf{\#Inst.} & \textbf{Capacity (KB)} & \textbf{\#Iter.}\\
		\midrule
		\multirow{5}{*}{\centering \textbf{AES-128}} & BitSlicing() & 288 & 0.58 & 2\\
            & AddRoundKey() & 24 & 0.05 & 11\\
		& SubBytes() & 357 & 0.71 & 10\\
            & ShiftRows() & 456 & 0.91 & 10\\
            & MixColumns() & 258 & 0.52 & 9\\
            \midrule
            \multirow{3}{*}{\centering \textbf{GHASH}} & ByteArrange() & 63 & 0.13 & 1\\
            & ByteAligning() & 138 & 0.28 & 8\\
            & GaloisMult() & 16 & 0.03 & 1024\\
            \midrule
            \textbf{SHA3} & StatePermute() & 633 & 1.27 & 24\\
            \midrule
             & \textbf{Total} & 2233 & 4.47 &\\
		\bottomrule
	\end{tabular}}
	\label{ctrl-overhead}
\end{table}

\subsection{Hardware Overhead Analysis}

The area overhead of \textit{CryptoSRAM} is approximately only 2\% compared to conventional SRAM arrays without computational capabilities. This is attributed to two main factors:
1) \textbf{Low control overhead}: Our control logic is simplified, as detailed in Section \ref{sec:control}. Furthermore, the size of the CMD array is minimized to just over 4.47KB due to our high degree of command reuse.
2) \textbf{Simplistic computational logic}: Compared to other compute SRAM designs \cite{agaComputeCaches2017e}, \textit{CryptoSRAM} only incorporates a 1-bit shift on top of basic logical operations. It avoids the inclusion of complex peripheral circuits, such as dedicated adders and shifters.

\section{Related Work}\label{relatedwork}

\textbf{Microcontroller Architectures and IoT Applications:} Microcontrollers have become integral components in a variety of IoT applications, including image, audio, and temperature sensing \cite{maheepala2020low,turchet2020internet,vanaja2018iot}. Many designs have focused on enhancing security and performance through software and hardware optimization \cite{chien2016low,jayakumarEnergyawareMemoryMapping2017,pearson2019misconception,ferretti2019fog}. Notably, certain standards and protocols such as AES and SHA3 have been established for encryption and authentication in these MCU architectures \cite{sarker2020lightweight}. However, these conventional architectures face significant challenges in meeting the ever-increasing demand for throughput and energy efficiency \cite{aerabi2020design}.

\textbf{Hardware Security Engines:} Existing research has proposed using hardware security engines for offloading encryption and authentication tasks from the CPU \cite{yangHardwareDesignsSecurity2017,yuLightweightMaskedAES2017,sravaniEfficiencyEnhancementSHA32022,luBPUBlockchainProcessing2020}. While these hardware engines have proven beneficial, they come with inherent limitations, such as performance bottlenecks, high energy consumption, and dependence on DMA \cite{munozAnalyzingResourceUtilization2018,biasizzoHardwareImplementationAES2005,contiIoTEndpointSystemonchip2017}. This necessitates an exploration of alternative solutions to overcome these challenges.

\textbf{In-SRAM Computing:} In-SRAM computing offers high parallelism and energy efficiency, promising solutions to the performance and power issues \cite{zhangSealerInSRAMAES2022a,zhangInhaleEnablingHighPerformance2022c,reisIMCRYPTOInMemoryComputing2022c}. Nevertheless, these techniques have largely been confined to non-MCU systems, where they incur significant overheads due to the need for ordered data placement within the SRAM subarrays \cite{zhangRecryptorReconfigurableCryptographic2018e,agaComputeCaches2017e,eckertNeuralCacheBitSerial2018a,fujikiDualityCacheData2019g}. Furthermore, the use of virtual addressing and set-associative cache complicates the application of in-SRAM computing in these systems.
Our proposed system integrates in-SRAM computing into MCU architectures, effectively addressing the limitations faced by the current solutions. Besides, by utilizing in-SRAM computing, our proposed system can support other applications, such as deep neural network (DNN) \cite{hsuAIEdgeDevices2019}. 

\section{Conclusion}\label{conclusion}
We highlight the advantageous characteristics of Microcontroller Units (MCUs), especially their usage of physical addressing and Direct Memory Access (DMA) management of SRAM, which are well-suited for in-SRAM computing. We propose the MCU-ISC, a low-overhead design, to accelerate cryptographic workloads. By employing techniques like bit slicing and \textit{lane-per-row}, we were able to leverage in-SRAM computing to significantly enhance secure data transmission in MCU systems. Our evaluation shows that the proposed MCU-ISC design outperforms conventional CPU and ASIC-based solutions in terms of throughput, energy efficiency.
These results underline the potential of our design to pave the way for more energy-efficient and high-performance secure data transmissions in future MCU systems. 


\section*{Acknowledgment}
This work is funded, in part, by NSF Career Award \#2339317, NSF \#2235398, the initial complement fund from UCR, and the Hellman Fellowship from the University of California.

\bibliographystyle{IEEEtran}
\bibliography{1-ref,paperpile}

\end{document}